\documentclass[a4paper,11pt]{article}
\pdfoutput=1 
\usepackage{aas_macros}
\usepackage{jcappub} 

\usepackage[T1]{fontenc} 
\usepackage[dvipsnames]{xcolor}
\usepackage{subcaption}
\usepackage{xspace}
\usepackage[normalem]{ulem}
\usepackage{multirow}
\usepackage{caption}
\usepackage{subcaption}
\usepackage{footnote}

\newcommand{\lcdm}{\ensuremath{\Lambda\text{CDM}\xspace}}
\newcommand{\dlum}{\ensuremath{d_\mathrm{L}}}
\newcommand{\diff}{\ensuremath{\mathrm{d}}}
\newcommand{\vect}[1]{\ensuremath{\boldsymbol{#1}}} 
\newcommand{\mat}[1]{\ensuremath{\mathbf{#1}}} 
\newcommand{\Omo}{\ensuremath{\Omega_{\text{m0}}}}
\DeclareMathOperator{\cov}{\mathrm{cov}}

\usepackage{scalerel}
\usepackage{tikz}
\usetikzlibrary{svg.path}

\definecolor{orcidlogocol}{HTML}{A6CE39}
\tikzset{
  orcidlogo/.pic={
    \fill[orcidlogocol] svg{M256,128c0,70.7-57.3,128-128,128C57.3,256,0,198.7,0,128C0,57.3,57.3,0,128,0C198.7,0,256,57.3,256,128z};
    \fill[white] svg{M86.3,186.2H70.9V79.1h15.4v48.4V186.2z}
                 svg{M108.9,79.1h41.6c39.6,0,57,28.3,57,53.6c0,27.5-21.5,53.6-56.8,53.6h-41.8V79.1z M124.3,172.4h24.5c34.9,0,42.9-26.5,42.9-39.7c0-21.5-13.7-39.7-43.7-39.7h-23.7V172.4z}
                 svg{M88.7,56.8c0,5.5-4.5,10.1-10.1,10.1c-5.6,0-10.1-4.6-10.1-10.1c0-5.6,4.5-10.1,10.1-10.1C84.2,46.7,88.7,51.3,88.7,56.8z};
  }
}

\newcommand\orcidicon[1]{\href{https://orcid.org/#1}{\mbox{\scalerel*{
\begin{tikzpicture}[yscale=-1,transform shape]
\pic{orcidlogo};
\end{tikzpicture}
}{|}}}}

\title{
{How to use GP:} 
Effects of the mean function and hyperparameter selection on Gaussian Process regression
}

\author[a,b]{Seung-gyu Hwang\orcidicon{0000-0002-6355-6263},}
\author[a,b]{Benjamin L'Huillier\orcidicon{0000-0003-2934-6243},}
\author[c,d]{Ryan E. Keeley\orcidicon{0000-0002-0862-8789},}
\author[b]{M. James Jee,}
\author[c,e]{Arman Shafieloo\orcidicon{0000-0001-6815-0337}}

\affiliation[a]{Department of Physics and Astronomy, Sejong University, 
Seoul 05006, Korea}
\affiliation[b]{Department of Astronomy \& Space Science, Yonsei University,  Seoul 03722, Korea}
\affiliation[c]{Korea Astronomy and Space Science Institute, 
Daejeon 34055, Korea}
\affiliation[d]{Department of Physics, University of California Merced, 
CA 95343, USA}
\affiliation[e]{University of Science and Technology, 
Daejeon 34113, Korea}

\emailAdd{sghwang@yonsei.ac.kr}
\emailAdd{benjamin@sejong.ac.kr}
\emailAdd{rkeeley@ucmerced.edu}
\emailAdd{mkjee@yonsei.ac.kr}
\emailAdd{shafieloo@kasi.re.kr}

\abstract{Gaussian processes have been widely used in cosmology to reconstruct cosmological quantities in a model-independent way. 
However, the validity of the adopted mean function  and hyperparameters, and the dependence of the results on the choice have not been well explored.
In this paper, we study the effects of the underlying mean function and the hyperparameter selection on the reconstruction of the distance moduli from type Ia supernovae.
We show that the choice of an arbitrary mean function affects the reconstruction: a zero mean function leads to unphysical distance moduli and the best-fit \lcdm\ to biased reconstructions.
We propose to marginalize over a family of mean functions and over the hyperparameters to effectively remove their impact on the reconstructions. We further explore the validity and consistency of the results considering different kernel functions and show that our method is unbiased.}

\begin{document}
\maketitle
\flushbottom


\section{Introduction}
\label{sec:introduction}
The nature of the accelerating universe is one of the biggest questions in modern physics and cosmology. In a matter-dominated universe, the expansion speed of the Universe should gradually decrease, but observations suggest that the cosmic expansion is accelerated by some unknown component of the Universe, dark energy (DE), e.g. \cite{1998AJ....116.1009R,1999ApJ...517..565P}. 
 
The role of DE is to create a repulsive force against gravity. 
DE can be described as a perfect fluid with  equation of state (EoS) $w=p/\rho$, which is the ratio between pressure $p$ and density $\rho$ of the fluid.  
The simplest model is the presence of a positive constant in the Einstein field equations, which can be modeled as a fluid with $w=-1$. 

However, as observations become more and more precise, some tensions arise such as the Hubble tension between local measurements of the Hubble-Lemaître constant $H_0$ \cite{2018ApJ...861..126R,2021arXiv211204510R} and inferences of this parameter from high-redshift experiments \citep{2020A&A...641A...6P}. 
These tensions, in addition to our ignorance of the nature of DE,  may lead to modifications to the standard model. 
A possibility is that DE is dynamical, i.e., that the EoS $w$ is an evolving function of the redshift.
For a flat Friedmann–Lemaître–Robertson–Walker  metric with EoS $w(z)$, the expansion history $h(z)=H(z)/H_0$ is described by 
\begin{equation}
    \label{eq:hofz}
      h^2(z)=\Omega_\mathrm{m0}(1+z)^3 
      +(1-\Omega_\mathrm{m0})\exp{\left(3\int_{0}^{z} \frac{1+w(z')}{1+z'}dz'\right)},
\end{equation}
where $\Omega_\text{m0}$ is the matter  energy density parameter at the present time.

Two main approaches are carried out in the literature to test cosmology beyond the concordance model. 
On one hand, theoreticians are building new models, usually adding new degrees of freedom, then compare them to the data and obtain constraints on their parameters. 
While this approach can give tight constraints on the parameters, the results
may be biased towards the considered models, and may even be meaningless if the model is not a good description of the reality.
On the other hand, model-independent, data-driven approaches can be applied to infer cosmological quantities without assuming a model, and the resulting reconstructions are thus not biased by a particular model assumption. 
These methods can be more flexible, and thus allow the detection of features in the data that could be missed by traditional model-fitting approaches. 
However, they can be prone to overfitting and the interpretation of the results is not always straightforward. 

Among these model-independent methods, Gaussian process (GP) regression \cite{2006gpml.book.....R} has become particularly popular in the last decade or so. 
Refs.~\cite{2010PhRvD..82j3502H, 2010PhRvL.105x1302H, 2011PhRvD..84h3501H} introduced GP to analyze cosmological observations phenomenologically.
In addition, the flexibility of model-independent approaches enable the discovery of interesting trends. 
In a nutshell, GP modeling implicitely makes the assumption that the data and the theoretical function to be reconstructed are jointly Gaussian, and that their distribution can be fully described by a mean function and a covariance kernel, itself defined by its hyperparameters.
While a large amount of work has been devoted to the choice of the kernel, few studies have investigated the dependence on the choice of the mean function and hyperparameters (e.g., \cite{2010PhRvD..82j3502H,2012PhRvD..85l3530S}).
Most notably, most of the literature simply uses zero as the mean function, and fixes the hyperparameters to the value that maximize the marginal likelihood, claiming that zero mean function is a general case.
As we will show, the choice of this mean function is not necessarily well motivated, in particular for non-stationary functions such as cosmic distances or expansions.
We will also demonstrate that the choice of the hyperparameters is also dependent on that of the mean function.
Moreover, we will suggest a rigorous approach to attain a completely Bayesian analysis with the GPR.

In this study, we focus on several aspects of GP regression, in particular, the effect of the mean function and the choice of the hyperparameters. 
In Section \ref{sec:methdata}, we describe the method and the  mock data used in this study. 
Section \ref{sec:res} presents our main results on the factors that affect reconstruction results. 
We summarize our findings in Section~\ref{sec:ccl}.

\section{Method and Data}
\label{sec:methdata}

\subsection{Gaussian process regression}
\label{sec:GPR}
A GP is a collection of random variables, such that every finite collection of those random variables has a multivariate normal distribution. 
The distribution of $f$ at any finite number of points $\vect x = (x_1,\dots,x_N)$ is specified by a mean function $\mu(\mathrm{\vect{x}})$ and a covariance matrix $\mat{K}$,
\begin{align}
    \label{eq:GP1}
      f(\mathrm{\vect{x}})&\sim\mathcal{GP}(\mu(\mathrm{\vect {x}}), \mat{K}),
      \intertext{where the covariance matrix $\mat{K}$ is defined by}
      [\mat{K}]_{ij} & = k(x_i,x_j),
\end{align}
where $k(x,x')$ is the covariance kernel.

Specifically, the mean and covariance functions are
\begin{equation}
    \label{eq:MunCov}
    \begin{aligned}
          &\mu(\vect{x})=\mathbb{E}[f(\vect{x})],
          \\
          &k(x_i, x_j)=\mathbb{E}[(f(x_i)-\mu(x_i))(f(x_j)-\mu(x_j))].
    \end{aligned}
\end{equation}

A covariance function is a kernel function describing the correlation between the random variables.
While many shapes of kernels have been considered in the literature, in this study, we will focus on the square exponential (SE) kernel, defined as 
\begin{equation}
    \label{eq:kernel_SE}
        k(x, x')=\sigma_f^2\exp{\left[-\frac{(x-x')^2}{2l^2}\right]}
\end{equation}
The squared exponential kernel has two hyperparameters that control the correlation length ($l$) and magnitude ($\sigma_f$) of the fluctuations.
Section~\ref{sec:kernel} explores the role of the kernel. 
 
Gaussian process regression (GPR) is used in cosmology to reconstruct a desired quantity from a discrete number of observations \cite{2010PhRvD..82j3502H, 2010PhRvL.105x1302H, 2012PhRvD..85l3530S, 2012JCAP...06..036S, 2020MNRAS.491.3983K, 2020MNRAS.494..819L}. 
Two approaches can be used to perform a regression.

\subsubsection{Prior approach: forward-modeling}
{The first approach is to model the function to be reconstructed as a GP.
We will refer to this approach as the \emph{prior approach}, or \emph{untrained GP} \cite{2010PhRvD..82j3502H, 2010PhRvL.105x1302H, 2011PhRvD..84h3501H,  2018PhRvD..97l3501J,  2020MNRAS.491.3983K, 2020MNRAS.494..819L}.}
In essence, it consists in the following steps:
\begin{enumerate}
    \item Choose a mean function $\vect {\mu^*}$, a covariance kernel $k$, and its  hyperparameters and calculate the covariance matrix $\mat K_{**}=K(\vect{X^*}, \vect{X^*})$.
    \item Generate a collection of random functions $\vect {f^*}$ following the Gaussian distribution $\vect{f^*}\sim\mathcal{N}(\vect {\mu^*}, \mat K_{**})$.
    \item For each of these random functions $f^*$, evaluate the goodness-of-fit, for instance by calculating the $\chi^2$ for the samples $\vect {f^*}$ evaluated against the observation $\vect y$.
    \item Obtain the conditional distribution $\vect {f^*}|\vect y$ weighted by the goodness-of-fit.
\end{enumerate}
{In other words, we generate the samples $f^*$ following a GP, and individually calculate the goodness-of-fit for the observation.
The data covariance is not used to generate $f^*$, but is only used to evaluate the $\chi^2$.}
In practice, this approach can be computationally expensive, since we have to generate a large quantity of samples. 
However, it allows for forward-modeling and combining data sets \cite{2010PhRvD..82j3502H, 2010PhRvL.105x1302H, 2011PhRvD..84h3501H,  2018PhRvD..97l3501J,  2020MNRAS.491.3983K, 2020MNRAS.494..819L}.

\subsubsection{Posterior approach: GP regression}
Because of the computational cost of the prior approach, a more commonly used approach is to model the data and the function to be reconstructed as a GP, i.e., assume that they are jointly distributed and follow a Gaussian distribution with a chosen mean and covariance.
This approach will be referred to as the \emph{posterior approach}, or \emph{trained GP}  \cite{2012PhRvD..85l3530S,2012JCAP...06..036S}. 
In order to take into account the observational error in the process, the covariance matrix of the observation can be added to the kernel matrix,
\begin{equation}
\label{eq:jointGPR}
    \begin{bmatrix}
        \vect{y}\\\vect{f^*}
    \end{bmatrix}
    \mathtt{\sim}~\mathcal{N}\left(
        \begin{bmatrix}
            \vect{\mu}\\\vect{\mu^*}
        \end{bmatrix},
        \begin{bmatrix}
            K(\vect{X}, \vect{X}) + C & K(\vect{X}, \vect{X^*})\\
            K(\vect{X^*}, \vect{X}) & K(\vect{X^*}, \vect{X^*})
        \end{bmatrix}
    \right),
\end{equation}
where $\vect{y}$ is the observation, $\vect{f^*}$ is the reconstruction, $\vect{\mu^{(*)}}$ is the input mean function, $K$ is a kernel matrix given by $[K(\vect{X^{(*)}}, \vect{X^{(*)}})]_{ij}=k(x^{(*)}_i, x^{(*)}_j)$, $C$ is the covariance matrix of the observation, and * denotes the test points where we wish to reconstruct the function.
The reconstruction is done with the conditional distribution to model a physical quantity given the observation as a GP,
\begin{equation} 
\label{eq:conddist}
    \vect{f^*}|\vect{y}\sim\mathcal{N}\left(\overline{\vect{f^*}},
    \cov(\vect{f^*})\right).
\end{equation}

For a given input mean function and hyperparameters, the GPR is readily computed with the conditional distribution via the following matrix calculation, thanks to the matrix inversion lemma:
\begin{equation}
\label{eq:posterior}
    \begin{aligned}
        &\overline{\vect{f^*}}=\vect{\mu^*}+K(\vect{X^*}, \vect{X})[K(\vect{X}, \vect{X}) + C]^{-1}(\vect{y}-\vect{\mu}),
        \\
        &\cov(\vect{f^*})=K(\vect{X^*}, \vect{X^*})-K(\vect{X^*}, \vect{X})[K(\vect{X}, \vect{X}) + C]^{-1}K(\vect{X}, \vect{X^*}).
    \end{aligned}
\end{equation}

We note that in the posterior approach, it is not necessary to actually draw the individual samples $\vect {f^*}$, and if one is only interested in the distribution of \vect {f^*}, all the information is contained in \vect{\overline{f^*}} and $\cov(\vect{f^*})$, which can significantly speed up the computation.

In order to complete the regression, one needs to assess the goodness-of-fit of the reconstruction. 
From Bayes' theorem, the marginal likelihood of the conditional distribution, eq.~\eqref{eq:conddist}, is given by
\begin{equation}
\label{eq:likelihood_margoverf}
    p(\vect{y}|\vect{X}, \vect{\mu}, \mathcal{H})=\int{p(\vect{y}|\vect{f}, \vect{X})p(\vect{f}|\vect{X}, \vect{\mu}, \mathcal{H})\diff\vect{f}},
\end{equation}
where $\mathcal{H}$ is a set of hyperparameters of the kernel function.
It is important to note that the marginal likelihood is dependent on a choice of the input mean function, kernel function, and its hyperparameters.
In practice, for computational reasons, the logarithm of the marginal likelihood (LML) is often used.

\subsubsection{Reconstruction of the derivative}
GP can be used to reconstruct not only a function but also its derivative since the derivative of a GP is also a GP.
Assuming that the same kernel function is used for the GP, the kernel function for the derivative is 
\begin{equation}
    \label{eq:kernel_derivative}
        \mbox{cov}\left(f_i, \frac{\partial{f_j}}{\partial{x_j}}\right)=\frac{\partial{k(x_i, x_j)}}{\partial{x_j}}, \hspace{1cm} \mbox{cov}\left(\frac{\partial{f_i}}{\partial{x_i}}, \frac{\partial{f_j}}{\partial{x_j}}\right)=\frac{\partial{^2k(x_i, x_j)}}{\partial{x_i}\partial{x_j}}.
\end{equation}
With the covariance for the derivative, the joint distribution of the observation $\vect{y}$, the reconstruction $\vect{f^{*}}$, and its derivative $\vect{f^{*}}'$ are \cite{2012PhRvD..85l3530S},
\begin{equation}
\label{eq:jointdGPR}
    \begin{bmatrix}
        \vect{y}\\\vect{f^*}\\\vect{f^*}'
    \end{bmatrix}
    \mathtt{\sim}~\mathcal{N}\left(
        \begin{bmatrix}
            \vect{\mu}\\\vect{\mu^*}\\\vect{\mu^*}'
        \end{bmatrix},
        \begin{bmatrix}
            K(\vect{X}, \vect{X}) + C & K(\vect{X}, \vect{X^*}) & K'(\vect{X}, \vect{X^*})\\
            K(\vect{X}, \vect{X^*})^T & K(\vect{X^*}, \vect{X^*}) & K'(\vect{X^*}, \vect{X^*})\\
            K'(\vect{X}, \vect{X^*})^T & K'(\vect{X^*}, \vect{X^*}) & K''(\vect{X^*}, \vect{X^*})
        \end{bmatrix}
    \right),
\end{equation}
where
\begin{equation}
\label{eq:dkernel}
[K'(\vect{X}, \vect{X^*})]_{ij}=\frac{\partial{k(x_i, x_j^*)}}{\partial{x_j^*}}, \hspace{1cm} [K''(\vect{X^*}, \vect{X^*})]_{ij}=\frac{\partial^2{k(x_i^*, x_j^*)}}{\partial{x_i^*}\partial{x_j^*}}
\end{equation}

Similarly to eq.~\eqref{eq:posterior}, the conditional distribution for the derivative given the observation also follow the Gaussian distribution:
\begin{align}
    \begin{bmatrix}
    \vect f\\
    \vect f'
    \end{bmatrix}|\vect y \sim
    \mathcal{N}\left(
    \begin{bmatrix}
    \bar {\vect f}\\
    \bar {\vect  f'}
    \end{bmatrix};
    \begin{bmatrix}
    \mat A-\mat D \mat B^{-1} \mat D^T
    \end{bmatrix}
    \right),
\end{align}
where
 \begin{align}
\label{eq:dposterior}
    \begin{bmatrix}
    \bar{\vect f} \\
    \bar{\vect f'}
    \end{bmatrix} &= 
    \mat D \mat B^{-1} 
    \begin{bmatrix}
    \vect y-\vect{\mu}
    \end{bmatrix}, \\
    \mat A & = 
    \begin{bmatrix}
    \mat K(\vect X^*, \vect X^*) &  \mat K'(\vect X^*, \vect X^*)\\
     \mat K'(\vect X^*, \vect X^*)^T &  \mat K''(\vect X^*, \vect X^*)
    \end{bmatrix}
    \\
    \mat B & = 
    \mat K(\vect X, \vect X) + \mat  C
    \\
    \mat D^T & = 
    \begin{bmatrix}
    \mat K(\vect X, \vect X^*)& 
     \mat K'(\vect X, \vect X^*)
    \end{bmatrix}.
\end{align}
The shapes of \mat A, \mat B, and \mat D are respectively $2n_*\times 2n_*$, $n\times n$, and $n_*\times n$, where $n_*$ is the size of $\vec f*$  and $n$ the number of data points.


\subsection{Data}
We generate mock data based on the Pantheon compilation \cite{2018ApJ...859..101S} for our fiducial cosmology. 
This state-of-the-art SNIa compilation contains 1048 points up to $z\simeq 2.3$. 
We generate a Gaussian white noise using the Pantheon covariance matrix, and add it to the fiducial cosmology at the redshifts of the actual Pantheon data points. 
This effectively creates a ``Pantheon-like'' mock dataset with a Gaussian noise. 
Since we aim to test the performance of GPR, we choose as a fiducial cosmology a non-\lcdm\ dark energy with $\Omo=0.3$, in our case, the phenomenologically emergent dark energy (PEDE) model \cite{2019ApJ...883L...3L}.
This model has the same number of degrees of freedom as \lcdm, and has a phantom equation of state in the asymptotic past and transition to a cosmological constant ($w=-1$) in the asymptotic future.

We compare PEDE to a commonly used parameterization of the DE equation of state, the Chevallier-Polarski-Linder (CPL) \cite{2001IJMPD..10..213C,2003PhRvL..90i1301L} parameterization. 
The CPL parameterization  is a Taylor expansion of $w$ at $z=0$:
\begin{align}
    w(a) & = w_0 + w_a(1-a). 
\end{align}
This two-parameter model has been widely used in the literature to describe the low-redshift evolution of the equation of state.

\begin{figure}[t]
\centering 
    \includegraphics[width=.47\textwidth]{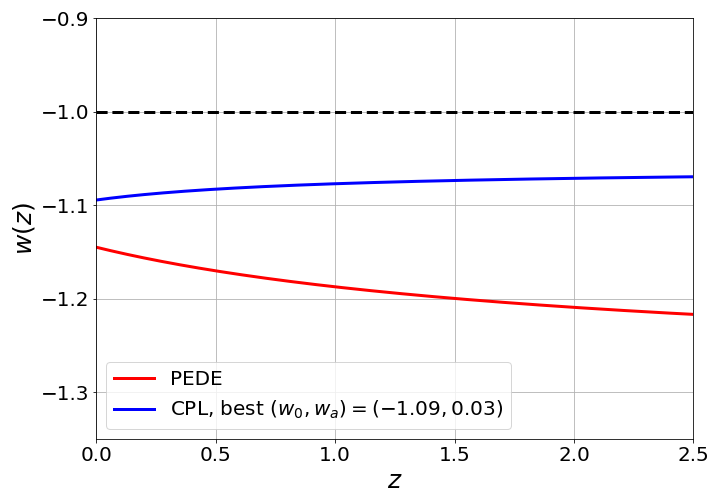}\hfill
    \includegraphics[width=.47\textwidth]{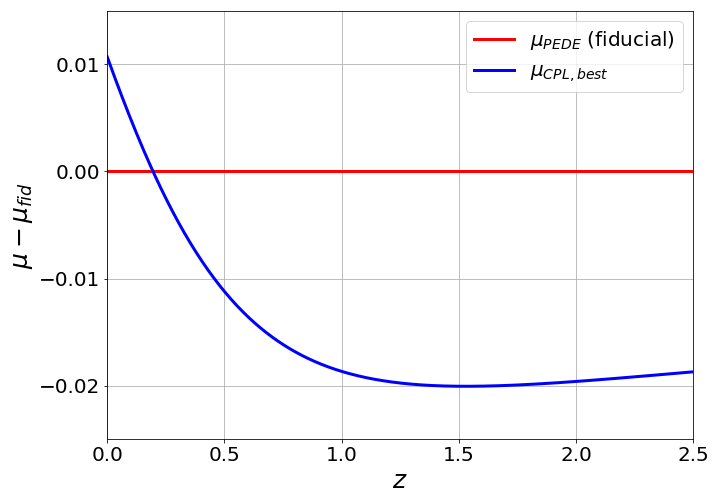}\hfill
    \caption{\label{fig:data} Fiducial cosmology for mock data; PEDE model with $(h, \Omega_m)=(0.7, 0.3)$. \emph{Left}: The equation of state of our fiducial cosmology and the CPL best-fit. \emph{Right}: Distance moduli of fiducial cosmology and the CPL best-fit.}
\end{figure}

Fig.~\ref{fig:data} shows the equation of state and the distance modulus for the fiducial (PEDE) model in red.
In blue is the best-fit for the CPL case with free parameters $(\Omega_m, w_0, w_a)$.
Although the distance moduli for both cases are similar (difference of about 0.02 mag at $z=1.5)$, the equation of states show different evolution.
Since $\mu$ is obtained by integrating twice, the difference between PEDE and the CPL best-fit is attenuated in $\mu$ space.
These results illustrate the fact that the fiducial cosmology is not included in the CPL parameterization.

\section{Results}
\label{sec:res}
In this section, we study in detail different aspects of GPR. 
Section~\ref{sec:meanfunc} shows the effect of the mean function, Section~\ref{sec:kernel} is devoted to the effect of the kernel, and  Section~\ref{sec:prior_vs_posterior} compare the prior and posterior approaches.

\subsection{Role of the mean function}
\label{sec:meanfunc}
Although the first applications of GPR to cosmology \cite{2010PhRvD..82j3502H,2010PhRvL.105x1302H,2011PhRvD..84h3501H,2012PhRvD..85l3530S} considered the effect of the mean function, many papers use an arbitrary mean function as an input such as a zero mean for convenience.
Refs.~\cite{2010PhRvD..82j3502H,2010PhRvL.105x1302H,2011PhRvD..84h3501H} used the prior approach and modeled the equation of state $w(z)$ and its first integral $y(z)=\int w(u)/(1+u)du$ as a GP, and fix the mean functions $\overline\mu(z)=-1$ and $\overline y(z) = -\ln(1+z)$, corresponding to the cosmological constant. 
Ref.~\cite{2012PhRvD..85l3530S} first applied an iterative smoothing method \cite{Shafieloo:2005nd,Shafieloo:2007cs,Shafieloo:2018gin} to prewhiten the data and obtain a mean function, then applied the posterior approach to reconstruct $\mu(z)$ directly from the data.
Because the iterative smoothing approach converges to a unique solution, the choice of the mean function is thus unique and defined by the data.
Refs.~\cite{2020MNRAS.491.3983K, 2021ApJ...922...95K} studied the LML distribution depending on the input mean function, 
and use the LML as a function of $\sigma_f$ to reject an input mean function.
Ref.~\cite{2022MNRAS.tmp..433R} used the $\lcdm$ best-fit with a multiplicative parameter as a mean function in order to give more flexibility than a constant mean function.  
While a constant mean function is legitimate in the case of stationary signals, in the case of a time-evolving quantity such as the distance moduli, or other cosmological quantities such as the Hubble expansion history, it is less clear.  
We will show that the zero mean cause a problem when reconstructing distance moduli and illustrate how a different choice of mean function yields very different posteriors.
In this section, we will use the posterior approach.

\subsubsection{Zero mean function}
\label{sec:zeromean}

\begin{figure}[t]
\centering 
    \includegraphics[width=.47\textwidth]{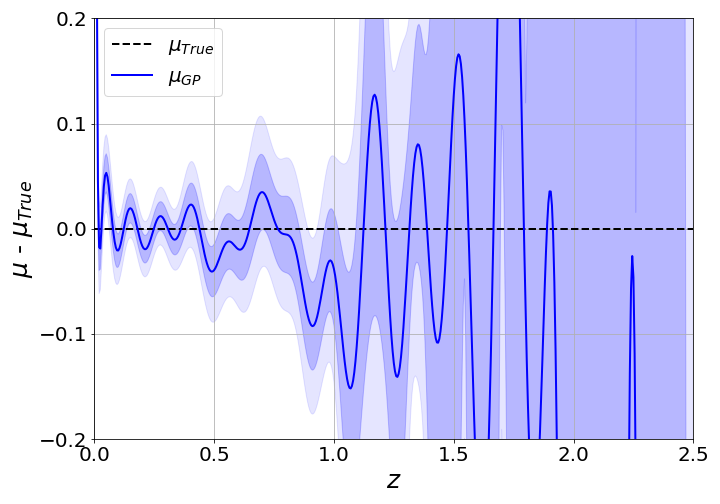}\hfill
    \includegraphics[width=.47\textwidth]{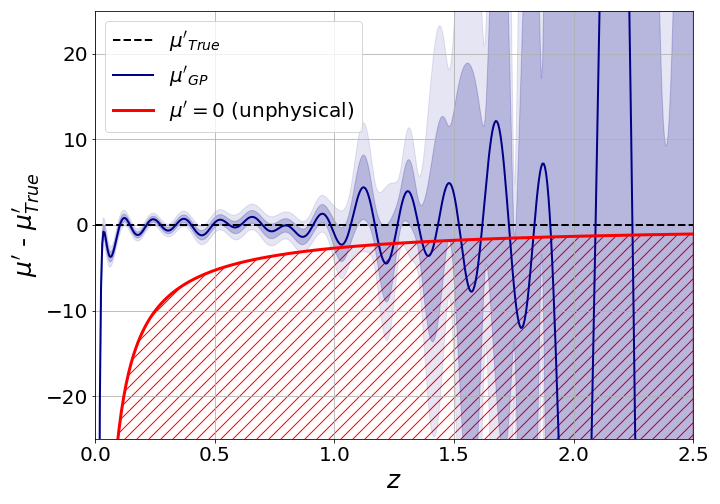}\hfill
    \caption{\label{fig:zero} GP regression and its derivative from \texttt{GaPP}. The blue solid line is the reconstructed mean of the GPR, and the blue bands show the 68\% and 95\% confidence levels. \emph{Left}: Reconstruction of the distance moduli $\mu$ with optimized hyperparameters assuming a zero mean function. \emph{Right}: GP regression for the derivative of the distance moduli $\mu'$ with \texttt{GaPP}. This regression is unphysical (below the red line, negative $\mu'$) beyond redshift $z\simeq1$.}
\end{figure}

Following Ref.~\cite{2012JCAP...06..036S}, a large number of studies in cosmology have used the \texttt{GaPP} package to reconstruct cosmological quantities. 
To illustrate our point, we will also use \texttt{GaPP} in this section. 
We verified that our own GP code and \texttt{GaPP} yield consistent results.
Negligible differences arise from the choice of the optimization algorithm.

The left-hand panel of Fig.~\ref{fig:zero} shows the difference between the predictive mean function and the true value to be reconstructed when using zero as the input mean function, as is done by \texttt{GaPP}. 
The best-fit hyperparameters for this realization are $(l, \sigma_f) = (0.13, 37.19)$.
While the reconstructed distance moduli are consistent with the truth, oscillations of the order of the best-fit hyperparameter $l$ are observed.

The right-hand panel of Fig.~\ref{fig:zero} shows the reconstructed derivative of the distance moduli. 
Oscillations are also visible in the derivative. This is understandable since the derivative involves a factor of order $l^{-2}$.  
In addition, beyond $z\simeq1$, the predictive mean becomes negative, which is unphysical, since 
\begin{align}
    \mu'(z) & = \frac 5 {\ln 10} \frac{\dlum'(z)}{\dlum(z)} >0. 
\end{align}
We conclude that zero is not a suitable choice of the  mean function to reconstruct the distance moduli.

\subsubsection{Best-fit as mean-function}
\label{sec:bfmean}
Next, we illustrate the effect of the mean function by using an alternative mean function.
We choose a mean function that we expect to describe the data relatively well, for instance the best-fit \lcdm\ case.

\begin{figure}
    \centering
    \begin{subfigure}[t]{0.47\textwidth}
    \includegraphics[width=\columnwidth]{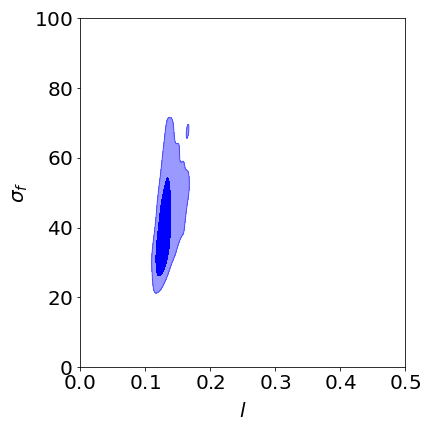}
    \caption{\label{fig:lml_zero} zero mean}
    \end{subfigure}\hfill 
    \begin{subfigure}[t]{0.47\textwidth}
    \includegraphics[width=\columnwidth]{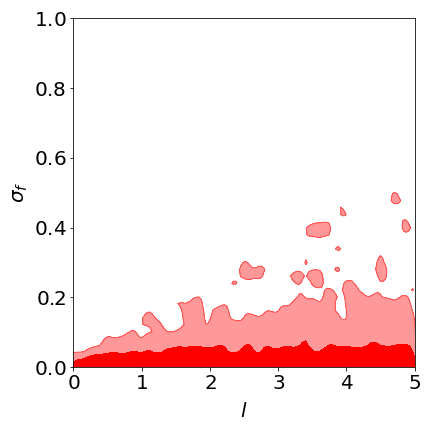}
    \caption{\label{fig:lml_bf} best-fit \lcdm\ mean}
    \end{subfigure}\hfill
    \caption{LML as a function of the hyperparameters obtained by MCMC sampling. \emph{Left}: The LML for the zero mean as the input mean function. \emph{Right}: The LML for the best-fit \lcdm\ mean as the input mean function.}
    \label{fig:lml_zero_bf}
\end{figure}

\begin{figure}
    \centering
    \begin{subfigure}[t]{0.47\textwidth}
    \includegraphics[width=\columnwidth]{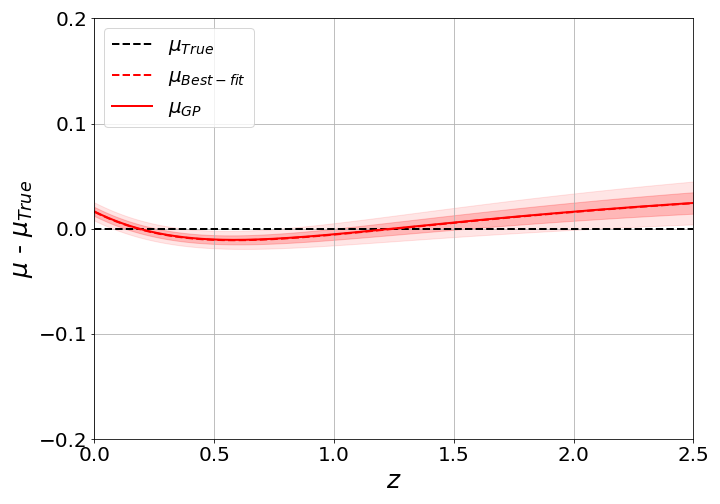}
    \caption{\label{fig:best_opt2} optimized hyperparameters}
    \end{subfigure}\hfill
    \begin{subfigure}[t]{0.47\textwidth}
    \includegraphics[width=\columnwidth]{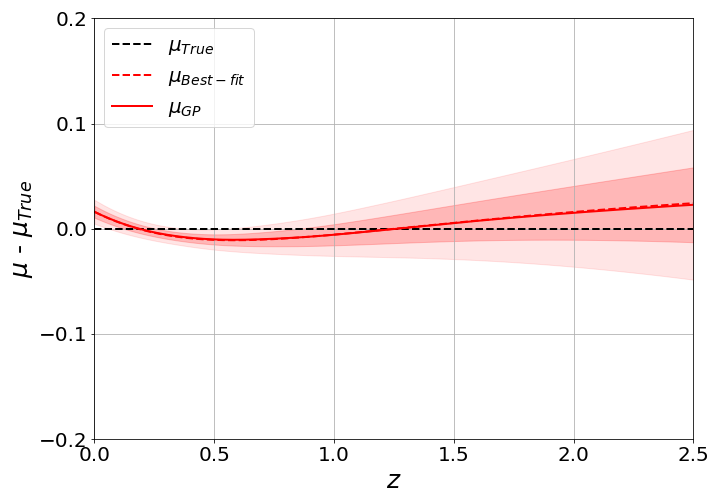}
    \caption{\label{fig:best_marg} marginalized hyperparameters}
    \end{subfigure}\vspace{0.03\textwidth}\hfill
    \begin{subfigure}[t]{.47\textwidth}
    \includegraphics[width=\columnwidth]{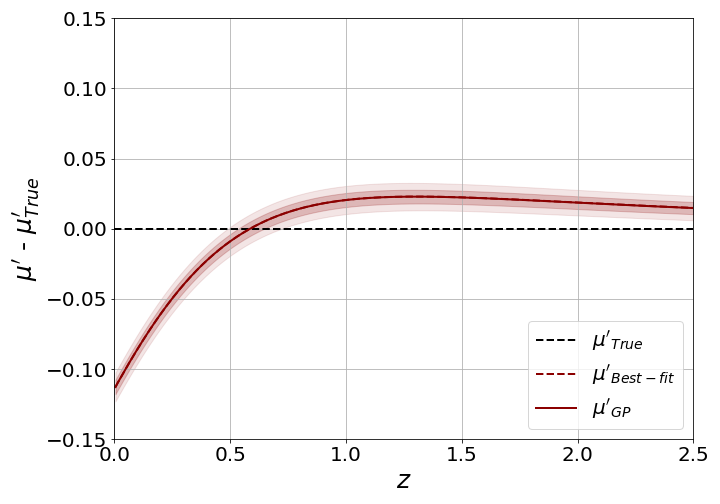}
    \caption{\label{fig:best_dmu_opt} optimized case: derivative of $\mu$}
    \end{subfigure}\hfill
    \begin{subfigure}[t]{.47\textwidth}
    \includegraphics[width=\columnwidth]{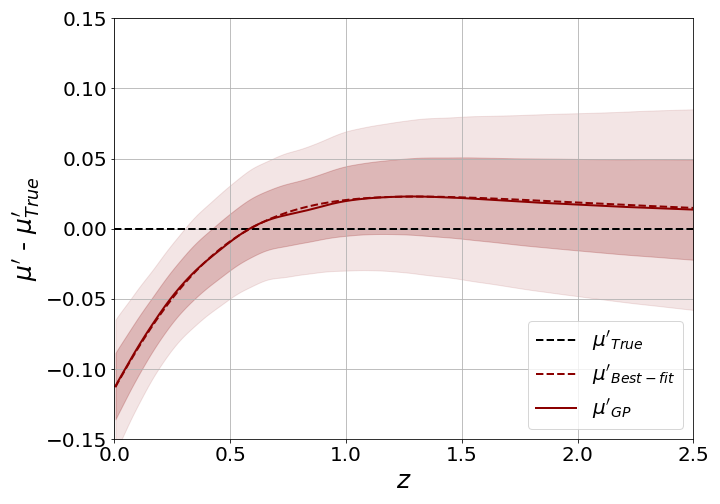}
    \caption{\label{fig:best_dmu_marg} marginalized case: derivative of $\mu$}
    \end{subfigure}\hfill
    \caption{GP regressions for different hyperparameter selection. \emph{Top-Left}: Optimized hyperparameters. \emph{Top-Right}: Marginalized hyperparameters. \emph{Bottoms}: Derivative of $\mu$ with optimized and marginalized hyperparameters. We stress out that the y-axis in this figure is 10 times smaller than Fig.~\ref{fig:zero}. 
    Therefore, the unphysical region $(\mu'<0)$ is not shown here.}
    \label{fig:best_hyper}
\end{figure}

\begin{table}[tbp]
\centering
\begin{tabular}{|c|c|cc|c|}
\hline
\multicolumn{2}{|c|}{} & $l$ & $\sigma_f$ & Note\\
\hline
\multirow{3}{4em}{Posterior approach} & Zero mean & $\mathcal U(0, 5)$ & $\mathcal U(0, 100)$ & Section~\ref{sec:bfmean} \\
& Best-fit mean & $\mathcal U(0, 5)$\footnotemark & $\mathcal U(0, 5)$ & Section~\ref{sec:bfmean} \\
& Full-marginalization & $\mathcal U(0, 5)$ & $\mathcal U(0, 5)$ & Section~\ref{sec:fullmarg} \\
\hline
{Prior approach} & Full-marginalization & $\mathcal U(0, 5)$ & $\mathcal U(0, 5)$ & Section~\ref{sec:prior_vs_posterior} \\
\hline 
\end{tabular}
\caption{\label{tab:prior_full} Priors for the hyperparameters and cosmological parameters. $\mathcal U$ denotes a uniform distribution.}
\end{table}
\footnotetext{{A different prior of $\mathcal U(0.05, 5)$ is used to reconstruct the derivative of distance modulus in Fig.~\ref{fig:best_dmu_opt} and~\ref{fig:best_dmu_marg}.
See Section~\ref{sec:bfmean}.}}

The effect of the mean function can be immediately seen in  Fig.~\ref{fig:lml_zero_bf}, which shows the LML as a function of the hyperparameters for both mean functions.
The LMLs are obtained by Markov Chain Monte Carlo (MCMC) algorithm as implemented in the \texttt{emcee} package \cite{2013PASP..125..306F}.
The priors for the hyperparameters are shown in Table~\ref{tab:prior_full}.
As noted by \cite{2012JCAP...06..036S}, in the zero mean-function case, the LML is sharply peaked at $(l, \sigma_f) = (0.13, 37.19)$, and therefore it can be well approximated by a Dirac distribution at this point. 
However, the situation is very different when using the best-fit \lcdm\ as an input mean function, as can be seen in the right-hand panel.
In this case, since the fit to the data is relatively correct, the LML is not sharply peaked and prefers a small value of $\sigma_f$. 
We stress the difference of scales in the $\sigma_f$ axis. 
In other words, since the input mean function is a ``good enough'' description of the data, GP does not find any preference for deviations, and the LML peaks at small values of $\sigma_f$.
This is clearly not the case in the zero mean function case, where the LML shows the need for a deviation from zero. 

On the other hand, there is no preference found for $l$ in the LML with the best-fit mean function as shown in Fig.~\ref{fig:lml_bf}.
The covariance of the derivative is roughly proportional to $l^{-2}$ when reconstructing the derivative.
Thus, if a small $l$ is chosen during MCMC sampling, its reconstruction of the derivative will have large error bars.
This will remain even after the hyperparameters are marginalized over.
To prevent too small $l$ being sampled, a prior with a lower bound for $l$ is chosen as commented on Table~\ref{tab:prior_full}.

Fig.~\ref{fig:best_opt2} and \ref{fig:best_dmu_opt} show the reconstructions of the distance moduli and its derivative using the optimal hyperparameters assuming the best-fit \lcdm\ as an input mean function.
As expected from the LML distribution in Fig.~\ref{fig:lml_bf}, the reconstruction is biased at less than $1\sigma$ towards the input mean function due to preference for small values of $\sigma_f$.

Since no sharp peak can be seen in the LML with the best-fit \lcdm\ mean,   many choices of hyperparameters can give reasonable fits to the data.
Therefore, in a Bayesian point of view, the meaningful approach is thus to marginalize the LML (eq.~\eqref{eq:likelihood_margoverf}) over the hyperparameter space.
\begin{equation}
\label{eq:likelihood_margoverfoverhypers}
    p(\vect{y}|\vect{X}, \vect{\mu})=\iint{p(\vect{y}|\vect{f}, \vect{X})p(\vect{f}|\vect{X}, \vect{\mu}, \mathcal{H})\diff\vect{f}\diff\mathcal{H}}.
\end{equation}
Fig.~\ref{fig:best_marg} and \ref{fig:best_dmu_marg} show the reconstructions for the marginalized case.
Although the reconstructed mean functions are consistent with the optimized case (Figs.~\ref{fig:best_opt2} and \ref{fig:best_dmu_opt}), the reconstructions with the optimized hyperparameters have narrower confidence levels.
These results show that the \emph{optimization} approach significantly underestimates the errors by not including other valid reconstructions, which are however included in the \emph{marginalization} approach.

\subsection{Full-marginalization}
\label{sec:fullmarg}

\begin{figure}
    \centering
    \includegraphics[width=0.9\textwidth]{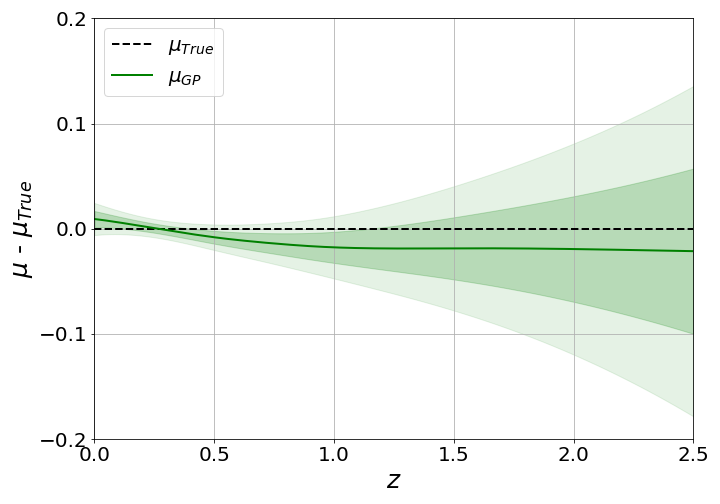}
    \caption{Full-marginalization with the CPL parameterization. The result is marginalized over five parameters; two hyperparameters $(l, \sigma_f)$ and three cosmological parameters $(\Omega_m, w_0, w_a)$.}
    \label{fig:full_one}
\end{figure}

In this section, we will introduce a rigorous application of GPR within a full Bayesian formalism.
As we saw previously, marginalizing the LML over the hyperparameter space is a natural Bayesian concept.
However, as we saw in Section~\ref{sec:meanfunc}, the choice of the mean function is also not well justified.
Therefore, we suggest full-marginalization, which is done by integrating the LML over the space of mean functions as well as hyperparameters:
\begin{equation}
\label{eq:likelihood_full}
    p(\vect{y}|\vect{X})=\iiint{p(\vect{y}|\vect{f}, \vect{X})p(\vect{f}|\vect{X}, \vect{\mu}, \mathcal{H})d\vect{f}d\mathcal{H}}d\vect{\mu}.
\end{equation}
Of course, in practice it is impossible to marginalize over the infinite space of possible mean functions. 
Instead, we chose a \emph{reasonable} family of functions, the CPL parameterization, to marginalize over.
CPL is a reasonable model to marginalize over, since we expect it to describe the data quite well.
However, as we saw, the fiducial model is not included in the CPL family.

We used the \texttt{emcee} package to marginalize over the cosmological parameters $(\Omo, w_0,w_a)$ and hyperparameters.
This idea is similar to the recent approach by Ref.~\cite{2022MNRAS.tmp..433R}, who use the best-fit \lcdm\ as a mean function and marginalize over a multiplicative parameter.

Fig.~\ref{fig:full_one} shows the result with full-marginalization for one realization.
Since the reconstruction with model-independent approach relies on a particular random realization, the result of full-marginalization has slight deviations from the truth at low redshift. We will discuss the bias of the methods we performed in section~\ref{sec:bias}.
Appendix~\ref{sec:realz} shows a hundred realizations.

\subsubsection{Role of the kernel function}
\label{sec:kernel}

\begin{figure}
    \centering
    \begin{subfigure}[t]{0.47\textwidth}
    \includegraphics[width=\columnwidth]{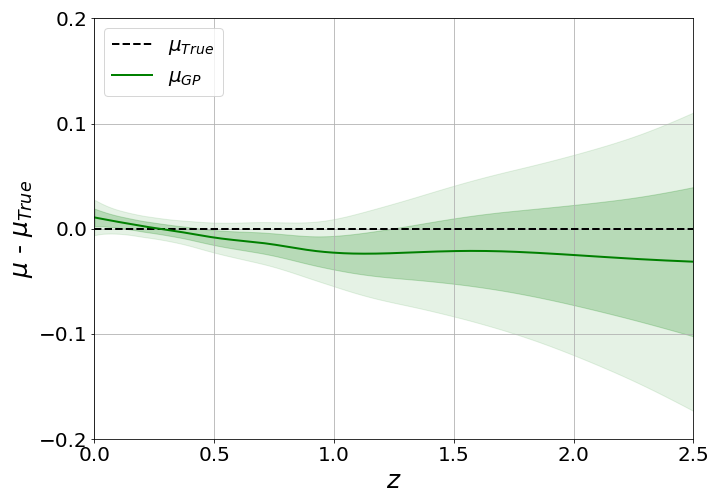}
    \caption{\label{fig:Matern32} Mat\'{e}rn kernel for $\nu=3/2$}
    \end{subfigure}\hfill
    \begin{subfigure}[t]{0.47\textwidth}
    \includegraphics[width=\columnwidth]{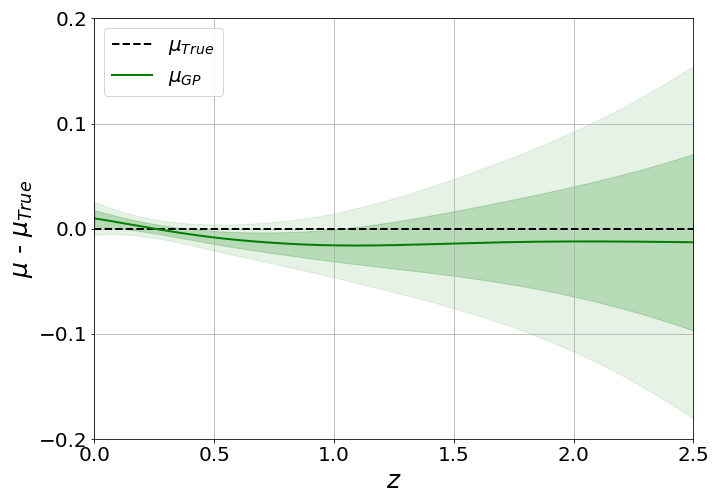}
    \caption{\label{fig:Matern92} Mat\'{e}rn kernel for $\nu=9/2$}
    \end{subfigure}\hfill
    \caption{Full-marginalization results with different kernels in the Mat\'{e}rn class.}
    \label{fig:diff_kernel}
\end{figure}

In this section, we test the role of the kernel function. 
Another common kernel is the Mat\'{e}rn class of kernel functions given by
\begin{equation}
    \label{eq:kernel_Matern}
        k(x, x')=\sigma_f^2\frac{2^{1-\nu}}{\Gamma\left(\nu\right)}\left(\frac{\sqrt{2\nu}\left(x-x'\right)}{l}\right)^{\nu}K_{\nu}\left(\frac{\sqrt{2\nu}\left(x-x'\right)}{l}\right)
\end{equation}
where $\nu>0$ is the smoothness parameter, $l$ and $\sigma_f$ are hyperparameters that play similar roles to the squared exponential case.
GP with the Mat\'{e}rn kernels allows to reconstruct up to the $n$th derivative of a function for $\nu>n$.
The $\nu\rightarrow\infty$ limit is the squared exponential function.
Ref.~\cite{2013arXiv1311.6678S} argued for the use of the Matérn kernels to reconstruct the equation of state, the comoving distance and its derivative on mock data, as the squared exponential kernel may be deemed too smooth.
Fig.~\ref{fig:diff_kernel} shows reconstructions with Mat\'{e}rn kernels for $\nu=3/2$ and $9/2$.
The reconstructions are essentially indistinguishable from the SE case, further demonstrating the importance of the mean function rather than the kernel on the reconstruction. 
The $\nu=3/2$ case shows more oscillations, as expected from the less smooth kernel.

\subsubsection{Bias of the method}
\label{sec:bias}

\begin{figure}
    \centering
    \begin{subfigure}[t]{0.47\textwidth}
    \includegraphics[width=\columnwidth]{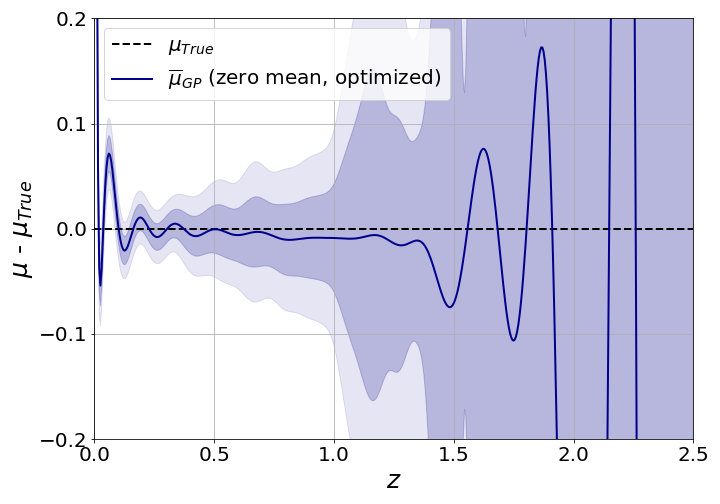}
    \caption{zero mean, optimized hyperparameters}
    \end{subfigure}\hfill
    \begin{subfigure}[t]{0.47\textwidth}
    \includegraphics[width=\columnwidth]{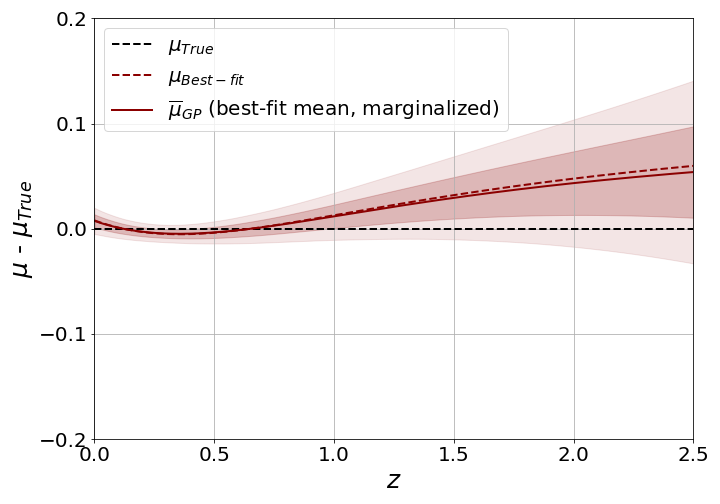}
    \caption{best-fit \lcdm\ mean, marginalized hyperparameters}
    \end{subfigure}\hfill
    \begin{subfigure}[t]{0.7\textwidth}
    \includegraphics[width=\columnwidth]{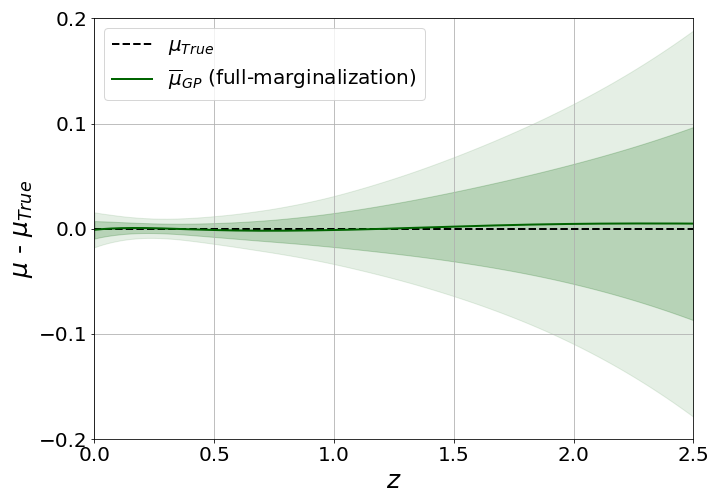}
    \caption{full-marginalization}
    \end{subfigure}
    \caption{Averaged results for 500 realizations for three different cases. \emph{Top-left}: Zero mean function, optimized hyperparameters. \emph{Top-right}: The best-fit \lcdm\ mean function, marginalized hyperparameters.
    \emph{Bottom}: Full-marginalization with the CPL parameterization.}
    \label{fig:superimpose}
\end{figure}

Fig.~\ref{fig:superimpose} shows the average of 500 realizations for three different choices of mean functions and hyperparameter selection. 
The top-left panel shows the averaged result for the zero-mean function with optimiziation of the hyperparameters (i.e., corresponding to the \texttt{GaPP} approach).
Clearly, unphysical oscillations are present at $z\geq 1$, where the data become scarce. 
As seen in Section~\ref{sec:zeromean}, this leads to an unphysical derivative. 
The top-right panel shows the case where the mean function is the best-fit \lcdm\ and we marginalized over the hyperparameters. 
While the behaviour is smoother than the zero-mean case, there is a small bias towards the input mean function (in red dashedline).
The bottom panel shows our suggested improvement, by marginalizing over the hyperparameters and a family of mean functions, in this case CPL. 
Even though the fiducial distance moduli (PEDE model) are not included in the family of input mean functions, the posteriors clearly show that our reconstructions are unbiased.

\subsection{Prior versus Posterior approach}
\label{sec:prior_vs_posterior}

\begin{figure}
    \centering
    \begin{subfigure}[t]{0.47\textwidth}
    \includegraphics[width=\columnwidth]{figure/fig04b_posterior_best_marg.png}
    \caption{\label{fig:posterior_marg} posterior approach}
    \end{subfigure}\hfill
    \begin{subfigure}[t]{0.47\textwidth}
    \includegraphics[width=\columnwidth]{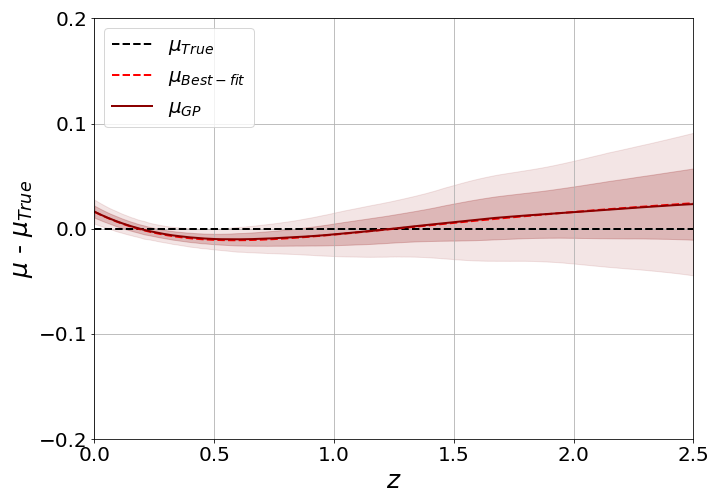}
    \caption{\label{fig:prior_marg} prior approach}
    \end{subfigure}\hfill
    \caption{Reconstruction of the distance moduli $\mu$ for marginalized hyperparameters with the posterior and prior approaches. \emph{Left}: Posterior approach. \emph{Right}: Prior approach.
    }
    \label{fig:prior_best}
\end{figure}

In this section, we compare the prior and posterior approaches.
We use the best-fit \lcdm\ mean function and marginalization over the hyperparameters as in the Section~\ref{sec:bfmean}.
Fig.~\ref{fig:prior_best} shows the reconstructions for the prior and posterior approaches.
Both cases give consistent behaviors, as expected mathematically.

However, the prior approach fails to reconstruct the distance moduli from a zero input mean function. 
Since the distance modulus increases rapidly from $z\simeq0$ and then slowly increases at high redshift, it is difficult for the GP to reconstruct the distance modulus. 
To be specific, reconstruction oscillates for the correlation length $(l)$. 
If a reconstruction increases sharply near $z\simeq0$, it should then decrease sharply again, which would give a bad fit to the distance moduli data. 
In other words, the reconstruction of the posterior approach with the zero input mean function is almost impossible in practice. 
Therefore, we suggest giving a reasonable input mean function for the GPR.

\section{Discussion and conclusion}
\label{sec:ccl}
We investigated the effects of the choice of the mean function, the hyperparameter selection in GP regression.
Our findings are summarized below. 
\begin{itemize}
    \item The choice of the mean function is important and should carefully be selected. Most of the literature so far fails to consider this, with the notable exceptions of \cite{2010PhRvL.105x1302H,2010PhRvD..82j3502H,2011PhRvD..84h3501H,2012PhRvD..85l3530S}. 
    \item Using a zero mean function on a nonstationary signal results in an artificial sharp peak in the LML, and by fixing the hyperparameters that maximize the LML, one miss many combinations that still give a good fit to the data. We advise against the use of zero as a mean function when the quantity is not stationary, as is the case for most cosmological quantities.
    \item We propose full-marginalization:  using a family of mean functions, i.e., non-flat $\Lambda$CDM or CPL, and marginalizing over the parameters associated with this mean function and the hyperparameters. This reduces the dependence on the input mean function. This approach is thus ``more'' model-independent, albeit still not exhaustive.
    \item Our method is unbiased, and allows to reconstruct the true mean function without any model assumption, as shown by Fig.~\ref{fig:superimpose}.
    \item We verified the equivalence between the ``prior'' and ``posterior'' approaches, which we expect from theory.
    The ``prior'' approach can be useful when fitting different data that are not linearly dependent, and for forward-modeling. 
\end{itemize}

We showed that arbitrarily setting the mean function to zero is not a neutral choice, and impact the hyperparameter selection. 
Fig.~\ref{fig:lml_zero_bf} shows that the observation by Ref.~\cite{2012JCAP...06..036S} that the marginal likelihood is sharply peaked, and that one may just use the best-fit hyperparameters, is actually an artifact of the choice of the mean function. 
If one chooses a mean function closer to the truth, such as the \lcdm\ best-fit as shown in Fig.~\ref{fig:lml_bf}, the LML peaks at low values of $\sigma_f$, independently of $l$ and the choice of the hyperparameters would not be so obvious. 
This is because GP is very efficient at picking deviations from the mean function. 
If the choice of the mean function is a close description of the data, the data show no preference for any deviations, and the LML peaks at $\sigma_f\simeq0$. 
Marginalizing over the hyperparameters becomes the obvious choice in the Bayesian point of view.

However, choosing the best-fit \lcdm\ model as the mean function also introduces a degree of arbitrariness, and the method is less model-independent. 
The reconstructions are biased towards the mean function, as shown in the top-right panel of Fig.~\ref{fig:superimpose}.

In practice, we suggest as an alternative to marginalize over a family of mean functions. 
We show the example of the CPL family, which yields unbiased reconstructions, despite the fact that the data were generated using a fiducial cosmology that is not included in CPL. 
As shown in Appendix~\ref{sec:dep_mean}, the results are independent of the family of mean functions, at least as long as a ``reasonable'' choice of mean functions is made. 

In this work, we marginalize over the CPL family of mean functions, while the data were generated using the PEDE model.
This is a reasonable choice, since CPL contains the concordance \lcdm\ model.  
Should the data be too different from a CPL, the choice of mean function should be adjusted accordingly. 
Appendix \ref{sec:dep_mean} shows that the method is independent of the choice of the family of mean functions.

Therefore, we recommend researchers in the field not to blindly use existing packages that assume a zero mean function and fix the value of the hyperparameters to the best-fit of the LML, but instead to use our full-marginalization approach. 
Given the popularity of GP in cosmology, we hope that the community will adopt these practices.
We note that in practice, \texttt{GaPP} has been applied to other quantities than the distance modulus, in particular, direct measurements of the expansion history via cosmic chronometers.

\acknowledgments
S.-g.~H. and B.~L. acknowledge the support of the National Research Foundation of Korea (NRF-2019R1I1A1A01063740). 
B.~L. and A.~S. acknowledge the support of the Korea Institute for Advanced Study (KIAS) grant funded by the government of Korea. A.~S. would like to acknowledge the support by National Research Foundation of Korea NRF-2021M3F7A1082053. 
 
\bibliographystyle{JHEP}
\bibliography{biblio}

\appendix
\newpage

\section{Dependence on the family of mean function}
\label{sec:dep_mean}

\begin{figure}
    \centering
    \begin{subfigure}[t]{0.47\textwidth}
    \includegraphics[width=\columnwidth]{figure/fig07c_avg_full_marg_CPL.png}
    \caption{\label{fig:full_avg_CPL} CPL parameterization $(\Omega_m, w_0, w_a)$}
    \end{subfigure}\hfill
    \begin{subfigure}[t]{0.47\textwidth}
    \includegraphics[width=\columnwidth]{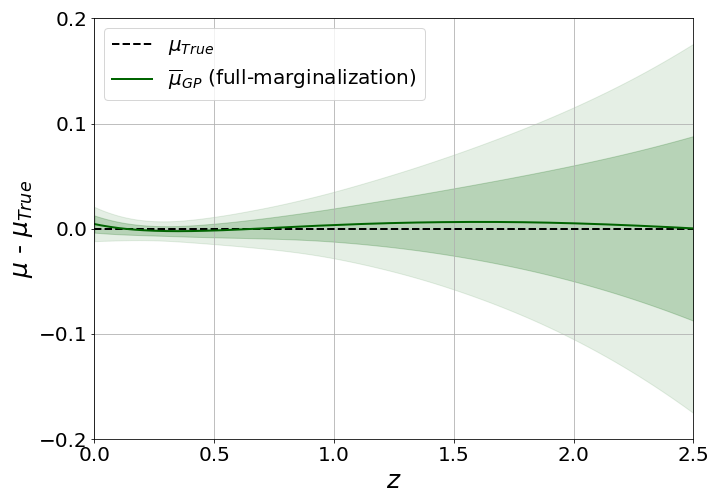}
    \caption{\label{fig:full_avg_oLCDM} non-flat \lcdm\ $(\Omega_m, \Omega_\Lambda)$}
    \end{subfigure}\hfill
    \caption{\label{fig:full_avg}Averaged full-marginalization results for two families of mean functions.}
\end{figure}

In this section, we compare the marginalization over the CPL family of mean functions to another family, in this case, non-flat \lcdm. 
The left and right hand panels of Fig.~\ref{fig:full_avg} respectively show the posteriors on the distance moduli when applying our full-marginalization approach using the CPL and non-flat \lcdm\ families of mean functions.
Clearly, the results do not depend on the choice of the family, as long as a reasonable choice of mean function is made.

\section{Individual realizations}
\label{sec:realz}

\begin{figure}
    \centering
    \includegraphics[width=\columnwidth]{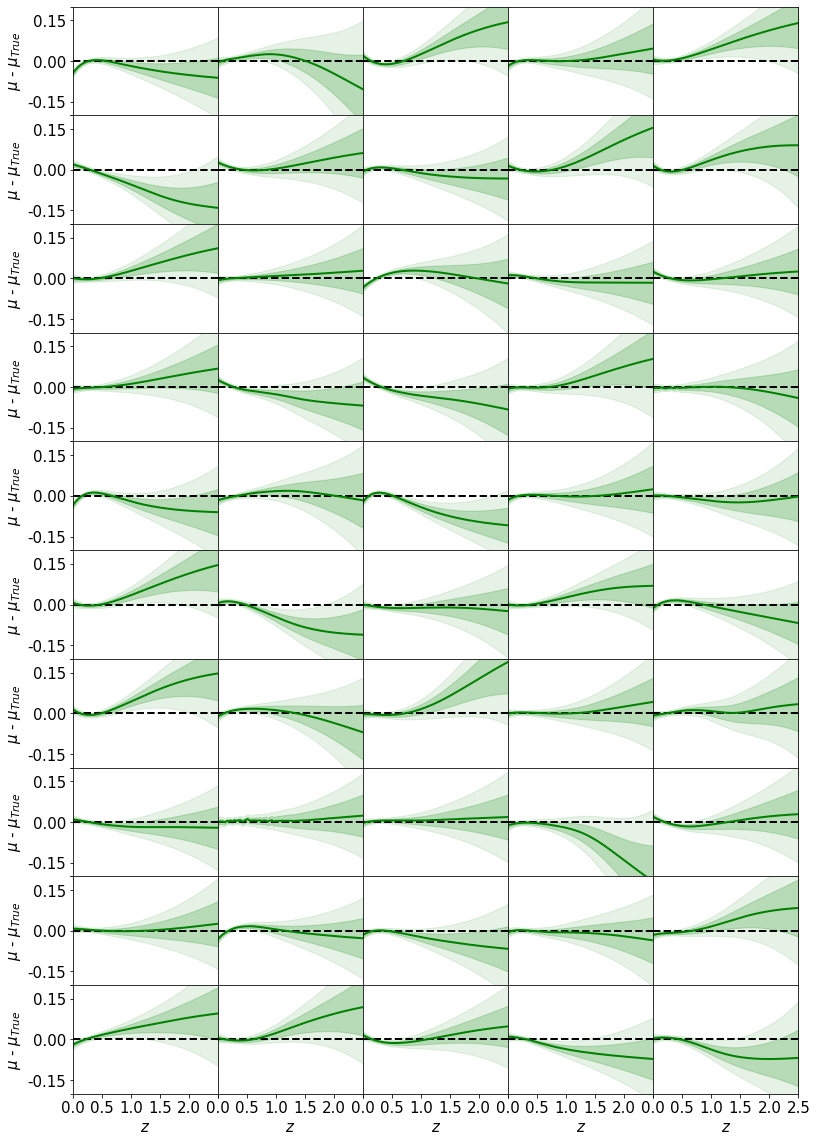}
    \caption{Full-marginalization for all realizations with the CPL parameterization.}
    \label{fig:all_CPL0}
\end{figure}

\begin{figure}
    \centering
    \includegraphics[width=\columnwidth]{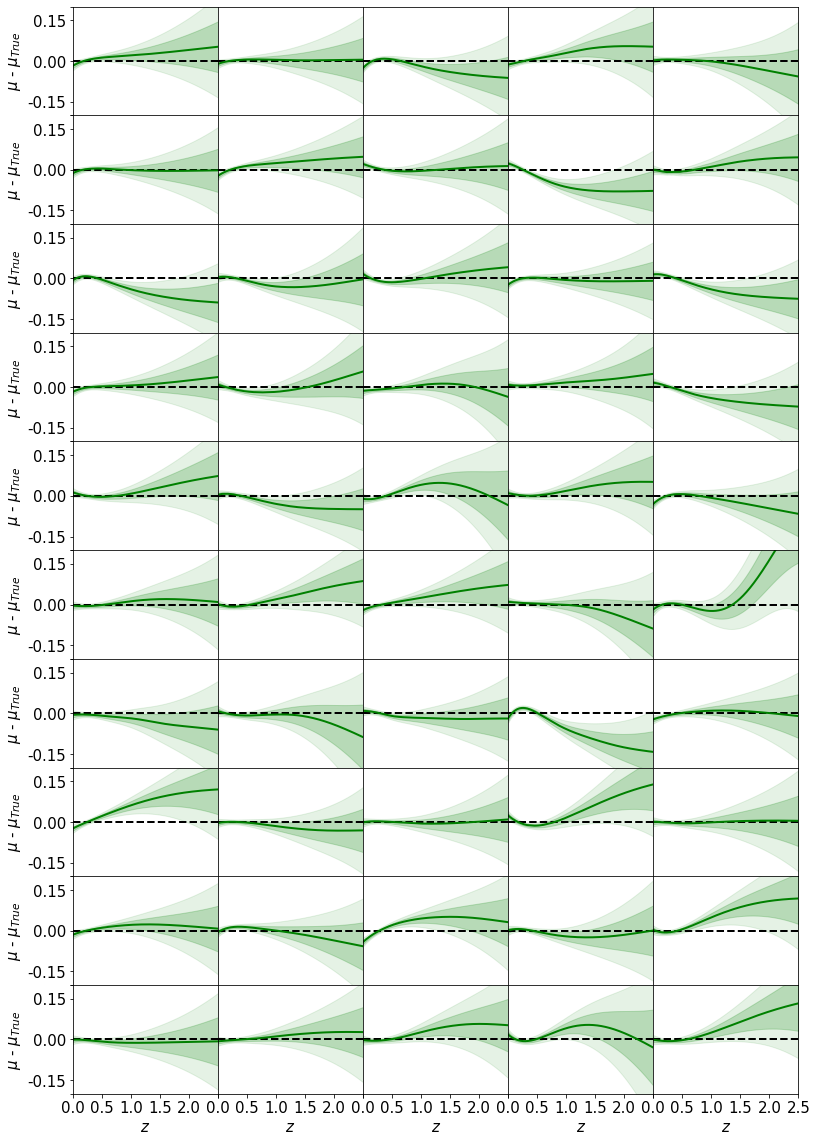}
    \caption{Full-marginalization for all realizations with the CPL parameterization.}
    \label{fig:all_CPL1}
\end{figure}

The figures presented in the body of the paper show the reconstructions for one particular realization of the mock data. 
Fig.~\ref{fig:all_CPL0} and \ref{fig:all_CPL1} shows the reconstructions for 100 of the 500 realizations used in the full marginalization case (Fig.~\ref{fig:superimpose}).

\end{document}